\documentclass[useAMS,usenatbib]{mn2e}

\usepackage{epsfig,times,deluxetable}

\def\MSUN   {\,$\hbox{M}_\odot$}

\def\kms    {km\,s$^{-1}$}
\def\apj    {ApJ}
\def\aap    {A\&A}
\def\apjs   {ApJS}
\def\procspie {Proc.~SPIE}
\def\mnras  {MNRAS}
\def\aaps   {A\&AS}
\def\araa   {ARA\&A}
\def\pasp   {PASP}
\def\apjl   {ApJL}
\def\pasj   {PASJ}
\def\aj     {AJ}

\title[Evidence for Large Grains in the Star-forming Filament
  OMC-2/3]{Evidence for Large Grains in the Star-forming Filament
  OMC-2/3}

\author[S. Schnee et al.]{Scott Schnee,$^{1}$\thanks{E-mail:
    sschnee@nrao.edu}, Brian Mason$^{1}$, James Di Francesco$^{2,3}$,
  Rachel Friesen$^{4,1}$, Di Li$^{5,6}$, \newauthor Sarah
  Sadavoy$^{7}$, and Thomas Stanke$^{8}$ \\
$^{1}$National Radio Astronomy Observatory, 520 Edgemont Road,
Charlottesville, VA 22903, USA \\
$^{2}$National Research Council Canada, Herzberg Institute of
Astrophysics, 5071 West Saanich Road Victoria, BC V9E 2E7, Canada \\
$^{3}$Department of Physics \& Astronomy, University of Victoria, PO Box
3055 STN CSC, Victoria, BC, V8W 3P6, Canada \\
$^{4}$Dunlap Institute for Astronomy and Astrophysics, University of
Toronto, 50 St. George Street, Toronto M5S 3H4, Ontario, Canada \\
$^{5}$National Astronomical Observatories, Chinese Academy of Science,
Chaoyang District Datun Rd A20, Beijing, China \\
$^{6}$Space Science Institute, Boulder, CO, USA \\
$^{7}$Max Planck Institute for Astronomy, Konigstuhl 17,
  D-69117 Heidelberg, Germany \\
$^{8}$ESO, Karl-Schwarzschild-Strasse 2, 85748 Garching bei
  M\"{u}nchen, Germany}
\begin{document}

\date{}

\maketitle

\begin{abstract}

We present a new 3.3\,mm continuum map of the OMC-2/3 region.  When
paired with previously published maps of 1.2\,mm continuum and
NH$_3$-derived temperature, we derive the emissivity spectral index of
dust emission in this region, tracking its changes across the filament
and cores.  We find that the median value of the emissivity spectral
index is 0.9, much shallower than previous estimates in other nearby
molecular clouds.  We find no significant difference between the
emissivity spectral index of dust in the OMC-2/3 filament and the
starless or protostellar cores.  Furthermore, the temperature and
emissivity spectral index, $\beta$, are anti-correlated at the
4\,$\sigma$ level.  The low values of the emissivity spectral index
found in OMC-2/3 can be explained by the presence of millimeter-sized
dust grains in the dense regions of the filaments to which these maps
are most sensitive.  Alternatively, a shallow dust emissivity spectral
index may indicate non-powerlaw spectral energy distributions,
significant free-free emission, or anomalous microwave emission.  We
discuss the possible implications of millimeter-sized dust grains
compared to the alternatives.

\end{abstract}

\begin{keywords}

ISM: dust -- ISM: Clouds -- stars: formation -- stars: protostars

\end{keywords}

\section{Introduction} \label{INTRODUCTION}

Dust emission is an excellent tracer of mass within star-forming
regions.  Unfortunately, deriving the mass from measurements of dust
continuum emission is not straightforward.  In particular, the
emissivities of the dust grains themselves are highly uncertain.
Generally, dust emissivity is assumed to have a power-law form at
millimeter wavelengths of $\kappa_\nu = \kappa_0(\nu/\nu_0)^\beta$,
and uncertainties in the emissivity spectral index, $\beta$, and the
relative dust opacity, $\kappa_0$, can result in derived masses that
are uncertain by factors of a few \citep[e.g.,][]{Ossenkopf94}.

The emissivity spectral index of dust is a difficult quantity to
measure.  For instance, the equation relating mass and the observed
emission (see Equation \ref{FLUXEQ} below) relies on the dust being
isothermal, which is not a valid assumption in any real star-forming
region.  Furthermore, when measuring dust emission near the peak of
the SED, variations in temperature along the line of sight and noise
in the continuum maps can make it difficult to derive accurate values
of $\beta$ \citep[e.g.,][]{Yang07, Shetty09a, Shetty09b, Schnee10,
  Ysard12}.  In particular, traditional $\chi^2$ minimization
techniques are susceptible to the aforementioned problems, whereas
Bayesian techniques can recover a more accurate value of $\beta$ from
emission maps near the peak of the dust SED \citep{Kelly12}.

The emissivity spectral index has been constrained observationally in
a few nearby molecular clouds and cores.  On small ($\sim$0.1\,pc)
scales, $\beta$ has been measured in L1448 \citep[$\beta = 2.44 \pm
  0.62$, 450\,\micron\ $\le \lambda \le$ 850\,\micron;][]{Shirley05}
and TMC-1C \citep[$\beta = 2.2 \pm 0.5$, 160\,\micron\ $\le \lambda
  \le$ 2100\,\micron;][]{Schnee10}.  Recent measurements of the
emissivity spectral index in small samples of Class 0 protostars find
$1.0 \le \beta \le 2.4$ \citep[862\,\micron\ $\le \lambda \le$
  3.3\,mm;][]{Shirley11} and $\beta \le 1$ \citep[1.3\,mm $\le \lambda
  \le$ 2.7\,mm;][]{Kwon09}.  The emissivity spectral index has also
been measured towards a few higher mass and more evolved objects, such
as hot molecular cores \citep[$\beta = 1.6 \pm 1.2$,
  800\,\micron\ $\le \lambda \le$ 870\,\micron;][]{Friesen05} and
ultracompact HII regions \citep[$\beta = 2.00 \pm 0.25$,
  60\,\micron\ $\le \lambda \le$ 1300\,\micron;][]{Hunter98}.  On
larger scales, i.e., those of filaments and molecular clouds, the
emissivity spectral index has been measured towards the Orion Nebula
Cluster \citep[$1.9 \le \beta \le 2.4$, 450\,\micron\ $\le \lambda
  \le$ 1100\,\micron;][]{Goldsmith97} and Perseus B1 \citep[$1.6 \le
  \beta \le 2.0$, 160\,\micron\ $\le \lambda \le$
  850\,\micron;][]{Sadavoy13}, both of which exhibit a flatter
spectral index towards the densest regions in the clouds.  On the
scale of an entire galaxy, \citet{Tabatabaei13} used measurements
between 70\,\micron\ $\le \lambda \le$ 500\,\micron\ to find that the
emissivity spectral index varies from $\beta \simeq 2$ in the center
of M33 to $\beta \simeq 1$ towards the edge of the galaxy.

The emissivity spectral index of dust measured in the laboratory
setting has a similarly large range ($1.5 \loa \beta \loa 2.5$), with
some materials showing a temperature-dependent value of $\beta$
\citep[e.g.,][]{Agladze96, Mennella98, Boudet05}.  The emissivity
spectral index is claimed to be anticorrelated with temperature in the
Orion, $\rho$ Ophiuchus and Taurus molecular clouds
\citep[e.g.,][]{Dupac03, PlanckAnticorrelation, Arab12}, though this
may be a result of line-of-sight temperature variations and
uncertainties in the observed fluxes rather than representing a true
property of the dust grains \citep{Shetty09a}.

The OMC-2/3 region, at a distance of 414\,pc \citep{Menten07, Kim08},
is the richest star-forming filament within 500\,pc.  Accordingly, it
has been studied intensively with instruments like SCUBA
\citep{Johnstone99, Nutter07}, MAMBO \citep{Davis09}, and Spitzer
\citep{Peterson05}.  These datasets have mapped the extent of the dust
emission in OMC-2/3, found dense cores within the filament, and
determined which cores harbour protostars \citep{Sadavoy10}.

In this paper, we present new 3\,mm continuum observations of the
Orion Molecular Cloud (OMC) 2/3 region.  We combine our new MUSTANG
3\,mm observations with a previously published 1.2\,mm map
\citep{Davis09} and a previously published gas temperature map
\citep{Li13} to derive the emissivity spectral index of dust in the
filaments and cores of the OMC-2/3 region.  The long wavelengths in
this analysis make the derived spectral index much less dependent on
the dust temperature than analyses of data at wavelengths around the
peak of the dust SED, substantially reducing a major source of
uncertainty in most other studies.

\section{Observations} \label{OBSERVATIONS}

Here we describe our new 3.3\,mm continuum map and the previously
published 1.2\,mm continuum and NH$_3$ maps of OMC-2/3.

\subsection{MUSTANG Observations} \label{NEWOBS}

The 3.3\,mm continuum emission from OMC-2/3 was mapped with the Robert
C. Byrd Green Bank Telescope (GBT) telescope in Green Bank, West
Virginia, using the MUSTANG array \citep{Dicker08}.  Data were
acquired in three observing sessions from November 2010 to February
2012 and were reduced and calibrated using procedures described as
outlined in \citet{Mason10}.  Data were visually inspected and $\sim
15\%$ were rejected, mostly due to poor weather. After flagging
problematic data, there were approximately 14 hours of integration
time spent on source. Flux calibration was accomplished by
periodically measuring the flux density of $\alpha$~Ori (Betelgeuse),
which was determined with reference to Uranus assuming $T=120 \pm 4 \,
{\rm K}$ \citep{Weiland11}.

A key step in the data reduction is the so-called ``common mode''
subtraction, consisting of subtracting an average of all good
detectors' signals at a given point in time from each detector's
signal. This is necessary to remove signals due to low-level
fluctuations in atmospheric emission, but has the side effect of
removing astronomical signals larger than approximately the MUSTANG
camera's instantaneous field of view.  A key difference between the
data reduction procedures of \citet{Mason10} and those here is that
the emission in OMC-2/3 is considerably brighter than the galaxy
cluster Sunyaev-Zel'dovich Effect being studied in the former work. It
is therefore feasible to reduce the amount of flux lost in extended
sources arising from common mode subtraction by iterating between the
map and timestream model domains, using the map estimated at the
previous step to correct the timestream (common mode) model, as
described in \cite{Dicker09}.  Spatial scales up to 2.4 arcminutes are
recovered by the reduction and calibration procedures, and the noise
in the resultant map is $\sim$0.36\,mJy/beam.

The GBT 3\,mm beam can be characterized by a Gaussian main beam
component (full-width at half-maximum (FWHM) of 8.6\arcsec\ and an
amplitude of 0.94) and a Gaussian error beam component (FWHM =
27.6\arcsec\ and amplitude of 0.06).  The relative contribution of the
GBT error beam on the 3\,mm map is roughly 20\%, with a maximum
possible of uncertainty of 30\% in some locations.  As a result, the
3\,mm map overestimates the flux found in an assumed 8.6\arcsec\ beam
by about 20\%, and we correct for this effect by multiplying our 3\,mm
maps by a factor of 0.8.  The resulting map, smoothed to 10.8\arcsec,
is shown in Figure~\ref{FLUXFIG}.

\subsection{MAMBO Observations} \label{MAMBO}

The 1.2\,mm continuum emission from OMC-2/3 was mapped with the IRAM
30\,m telescope at Pico Veleta, Spain, using the MAMBO array.  Data
were acquired between 1999 and 2002 and were reduced using the MOPSI
package following standard reduction steps.  See \citet{Davis09} for a
more detailed description of the observations and analysis.  The
spatial scales recovered in the MAMBO map are greater than those of
the MUSTANG map, a complexity that is discussed in Section
\ref{FILTERING}.  The noise in the MAMBO map is
$\sim$8\,mJy\,beam$^{-1}$.  The IRAM 30\,m beam has been characterized
at 1.3\,mm by a Gaussian main beam (FWHM = 10.5\arcsec\ and amplitude
0.975), a Gaussian first error beam (FWHM = 125\arcsec\ and amplitude
0.005) and Gaussian second and third error beams with larger FWHM and
lower amplitudes \citep{Greve98}.  Because the IRAM 30\,m error beams
have such small amplitudes relative to the main beam, we do not modify
the 1.2\,mm intensities as was done for the 3\,mm map.

The observed 1.2\,mm emission map is also shown in Figure
\ref{FLUXFIG}.  The 3\,mm and 1\,mm emission maps look qualitatively
very similar, as expected if thermal dust emission on the
Rayleigh-Jeans tail is the dominant emission mechanism.

\subsection{Ammonia Observations} \label{AMMONIA}

The NH$_3$ (1,1) and (2,2) lines were observed towards OMC-2/3 by the
Very Large Array (VLA) and the Green Bank Telescope (GBT).  Details of
the observations and calibration are given in \citet{Li13}.  The
combined VLA and GBT map has a spatial resolution of
$\sim$5\arcsec\ and a spectral resolution of 0.6\,\kms.  The kinetic
temperature of the gas was derived in \citet{Li13} following the
procedures outlined in \citet{Li03} and \citet{Ho83}.

In this paper, the temperature at each position is determined from a
smoothed and regridded $T_{gas}$ map from \citet{Li13}, putting it on
the same $\sim$11\arcsec\ resolution grid of the 1.2\,mm and 3.3\,mm
continuum maps.  Any position included in the continuum maps but not
included in the NH$_3$ maps was assigned the median temperature value
of the map (16.5\,K).  The range of temperatures in this analysis ($12
\leq T \leq 20$) is relatively narrow, so an assumed value of 16.5\,K
will not be too far off the true line-of-sight average temperature for
any region in OMC-2/3.  Roughly 4\% of the pixels in our analysis were
assigned the median temperature value of 16.5\,K.  For our analysis,
we assume that the gas and dust temperatues are coupled, a valid
assumption for densities greater than 10$^4$\,cm$^{-3}$ \citep{Doty97,
  Goldsmith01}.  The dust temperature map thus derived is shown in
Figure \ref{TEMPBETAFIG}.

\subsection{Spatial filtering} \label{FILTERING}

Ground-based bolometer maps all suffer from some degree of spatial
filtering as part of the data reduction process, to remove significant
atmospheric emission.  The MUSTANG 3\,mm map recovers less extended
emission than the MAMBO 1\,mm map, with spatial scales greater than
$\sim$2.4\arcmin\ filtered out.  We therefore pass the MAMBO data
through the the MUSTANG pipeline because the former have sensitivity
to emission on larger scales than the latter.  This operation ensures
that the same spatial filtering are applied to both the 1\,mm and the
3\,mm observations.  This technique is similar to that employed by
\citet{Sadavoy13} to compare {\it Herschel} and JCMT data.  As a
result of this filtering, the spectral indices determined in this
paper are derived for the smaller-scale material in the OMC-2/3
filament and cores and are not indicative of the more diffuse material
in the Orion molecular cloud.  In the relatively dense material traced
by the MUSTANG observations, the gas and dust temperatures are
expected to be well coupled (see Section \ref{AMMONIA}).  We smooth
the MUSTANG 3\,mm and NH$_3$ temperature observations to the MAMBO
resolution, then we place all three maps on a common grid.  We
consider only those pixels with signal to noise greater than 10 in
both the MAMBO map and the MUSTANG map.

\section{Analysis} \label{ANALYSIS}

We assume that the mm-wavelength emission observed along the
line-of-sight as a function of frequency is given by a modified
blackbody, i.e.

\begin{equation} \label{FLUXEQ}
S_\nu = \Omega B_\nu(T_d) \kappa_\nu \mu m_H N_{H_2}
\end{equation}
where
\begin{equation} \label{PLANCKEQ}
B_\nu(T_d) = \frac{2 h \nu^3}{c^2} \frac{1}{\exp(h \nu / k T_d) - 1}
\end{equation}
and 
\begin{equation} \label{KAPPAEQ}
\kappa_\nu = \kappa_{230} \left(\frac{\nu}{230~{\rm GHz}} \right)^\beta
\end{equation}  

In Equation \ref{FLUXEQ}, $S_\nu$ is the flux density per beam,
$\Omega$ is the solid angle of the beam, $B_\nu(T_d)$ is the blackbody
emission from the dust at temperature $T_d$, $\mu = 2.8$ is the mean
molecular weight of interstellar material in a molecular cloud per
hydrogen molecule, $m_H$ is the mass of the hydrogen atom, $N_{H_2}$
is the column density of hydrogen molecules, and a gas-to-dust ratio
of 100 is assumed.  In Equation (\ref{KAPPAEQ}), $\kappa_{230} =
0.009$ cm$^2$ g$^{-1}$ is the emissivity at 230\,GHz of the dust
grains at a gas density of 10$^6$~cm$^{-3}$ covered by a thin ice
mantle \citep[][Column 6 of Table 1]{Ossenkopf94} and $\beta$ is the
emissivity spectral index of the dust.

Equation \ref {FLUXEQ} assumes that the dust emission is optically
thin, which may be a concern towards the peak of the dust spectral
energy distribution (SED) but is of no concern for observations on the
Rayleigh-Jeans tail of the SED.  For instance, assuming standard dust
to gas ratios and dust properties \citep{Ossenkopf94}, the 1.2\,mm
emission from a dense core with a $V$-band extinction of $A_V = 100$
would have an optical depth of 0.004, while one would expect an
optical depth of 0.1 for 350\,\micron\ emission.  The assumption of
optically thin emission at 1\,mm and 3\,mm is justified for the
observations presented here.

The main assumptions required for Equation \ref{FLUXEQ} to reflect
accurately the emission as a function of frequency are 1) a lack of
free-free emission, 2) a lack of anomalous microwave emission, and 3)
a weak temperature dependence.  We discuss these different cases in
Section \ref{DISCUSSION}.

\subsection{Spectral index map} \label{BETAMAP}

The emissivity spectral index can be determined from the ratio of the
MAMBO and MUSTANG emission maps.  The spectral index at each position
in the map is given by:
\begin{equation} \label{BETAEQ}
\beta = \frac{{\rm ln}(S_{\rm 250\,GHz}/S_{\rm 90\,GHz})-{\rm
    ln}(B_{\rm 250\,GHz}(T_d)/B_{\rm 90\,GHz}(T_d))}{{\rm
    ln}{\rm (250\,GHz/90\,GHz)}}
\end{equation}
with $S_\nu$ and $B_\nu(T_d)$ as defined in Section \ref{ANALYSIS}.
We assume that the dust temperature is equal to the gas temperature
\citep[see Section \ref{RJ};][]{Li13}.  The resultant emissivity
spectral map is shown in Figure \ref{TEMPBETAFIG}.

\subsection{Getsources}

The OMC-2/3 region has a complex background of filamentary structures
that are not filtered-out in our data (see Figure \ref{FLUXFIG}). As
such, source extraction algorithms that use intensity maxima and
minima to identify structures \citep[e.g., CLUMPFIND;
  see][]{Williams94} will include emission from these filaments in
their extractions.  Therefore, we identified sources using the {\it
  getsources} algorithm, which includes a prescription for filament
extraction in addition to compact structures \citep[see ][for
  details]{Menschikov12}.

Briefly, {\it getsources} uses multiple spatial decompositions to
identify structure over many scales and characterize source masks,
which are defined by two dimensional Gaussians.  Information from the
source masks (and filament masks) are combined over each wavelength to
improve the final source footprint (i.e., in the case of blended
sources at lower resolution) and to extract source properties. For
efficiency, we masked out the edges of the 1\,mm and 3\,mm maps to
avoid noisy regions (and false detections) associated with the
low-coverage areas.

Using {\it getsources} version 1.140127, we identified initially 42
sources.  This initial catalogue may include unreliable detections,
e.g., due to the filtering.  Therefore, we required all sources to be
well detected in the single-scale decompositions (i.e., with
significance $>$ 1).  Of the initial 42 sources, we rejected 12 for
having either a low global significance or a low significances in
either the 1 mm or 3 mm maps.  Table \ref{SOURCETABLE} gives the
positions, sizes, and 3\,mm fluxes associated with the 30 robust
sources.

\section{Discussion} \label{DISCUSSION}

The median value of the emissivity spectral index between 1\,mm and
3\,mm in OMC-2/3 is $\beta = 0.9$, with a standard deviation of 0.3.
The typical uncertainty due to noise in the spectral index at any
given point in our map of $\beta$ is 0.16.  The systematic uncertainty
in $\beta$ due to absolute flux calibration uncertainties is 0.17.

The spectral energy distribution in OMC-2/3 is much shallower than has
been measured in OMC-1, where $1.9 \le \beta \le 2.4$  
\citep{Goldsmith97}.  In the following, we discuss possible
explanations for the low values of $\beta$ determined in OMC-2/3.

\subsection{Variations in $\beta$} \label{VARIATIONS}

The spectral indices ranged from $\beta \simeq 1.9$ to $\beta \simeq
-0.1$ (note that a blackbody would have $\beta = 0$).  This variation
is significantly larger than the uncertainties in $\beta$, so the
ratio of the 1\,mm flux to 3\,mm flux is strongly dependent on
position.  

We find no significant differences between the mean values of $\beta$
found towards starless cores, protostellar cores, or at positions
within the OMC-2/3 filament not associated with a core, with core
positions taken from \citet{Sadavoy10}.  The position of possible
protostars in OMC-2/3 (asterisks and diamonds in Figure
\ref{TEMPBETAFIG}), correspond to regions with the full range of steep
to shallow SEDs.  There are relatively few points associated with
starless cores and filamentary material removed from dense cores, but
these regions also have the same average spectral index and spread
($\beta \simeq 0.8-0.9 \pm 0.3$).  One starless core in particular,
J053530.0-045850 in the nomenclature of \citet{Sadavoy10}, is notable
for its low emissivity spectral index ($\beta \simeq 0.6$).  We
suggest that this core would be an interesting target for future
observations.

\subsection{Large grains} \label{LARGEGRAINS}

Values of emissivity spectral index with $\beta < 1$ have been seen
before in the disks around pre-main sequence stars and brown dwarfs
\citep[e.g.,][]{Lommen07, Perez12, Ricci12, Ubach12} and are
attributed to dust grain sizes up to millimeter scales.  Seven dense
cores in the Orion B9 region have had their spectral indices measured
between 350\,\micron\ $\le \lambda \le$ 850\,\micron, with a median
value of $\beta = 1.1$ \citep{Miettinen12}.  Large grains provide a
possible explanation for the relatively flat spectral index found in
the Orion cores in the \citet{Miettinen12} sample.  The presence of
millimeter-sized grains is also given as the likely explanation for
$\beta < 1$ found in three Class 0 protostellar cores in the 1.3\,mm -
2.7\,mm wavelength range by \citet{Kwon09}.

Similarly, the low values of $\beta$ in OMC-2/3 can be attributed to
much larger dust grains than are typically found in the diffuse ISM.
The MUSTANG and MAMBO maps do not recover large-scale features
associated with diffuse emission in OMC-2/3, so we have only measured
the emissivity spectral index of the more compact (i.e., likely the
densest) portions of the filaments as well as the dense starless and
protostellar cores.  Note that \citet{Johnstone99} find a typical
density for this filament of $\sim$10$^5$\,cm$^{-3}$, much lower than
is found in circumstellar disks but very similar to the densities
found by \citet{Miettinen12} in cores in Orion B9.  In OMC-1,
\citet{Goldsmith97} find that the spectral index decreases with
density, which is qualitatively consistent with the idea that the
observations presented here of the densest portions of OMC-2/3 are
characterized by a low average value of $\beta$.

Adopting the dust grain model presented in \citet{Ricci10}, we find
that a spectral index of $\beta \simeq 0.9$ indicates that the maximum
grain size is at least 1\,mm and could be as large is 1\,cm depending
on the powerlaw slope of the dust grain size distribution.  The dust
opacity at 1\,mm implied by the range of dust grain distributions and
maximum grain size is $4 \le \kappa_{1\,mm} \le 10$ cm$^2$\,g$^{-1}$
\citep{Ricci10}.  This range of $\kappa_{1\,mm}$ values is roughly a
factor of 3 - 7 larger than the standard value given
\citet{Ossenkopf94} Table 1 column 6.  

Core masses can be determined from the MUSTANG 3\,mm map using
\begin{equation} \label{MASSEQ}
M = \frac{D^2 S_{\rm 90\,GHz}}{\kappa_{\rm 90\,GHz} B_{\rm 90\,GHz}(T)}
\end{equation}
where D is the distance to OMC-2/3.  Since the mass derived from
Equation \ref{MASSEQ} varies inversely with the opacity, core masses
estimated from their 1\,mm flux would be a factor of 3 - 7 lower
assuming large grains than assuming standard \citet{Ossenkopf94}
grains.  Taking the median dust emissivity spectral index of $\beta =
0.9$, we find that core masses derived from 3\,mm emission are a
factor of 2.8 smaller than the masses determined with the commonly
assumed value of $\beta = 2$.  The mass derived from 3mm emission,
combining the effects of a higher opacity and shallower emissivity
spectral index, would be a factor of 10 - 20 lower for a model with
large grains and $\beta$ = 0.9 than one would derive from an
extrapolation of the \citet{Ossenkopf94} value at 1\,mm using and
emissivity spectral index of 2.  For example, a typical core in
OMC-2/3 has an integrated flux of 20\,mJy (see Table
\ref{SOURCETABLE}) at 3\,mm.  The mass derived for this typical core,
assuming a temperature of 16.5\,K (see Section \ref{AMMONIA}), would
be 2.8\,\MSUN\ for $\beta = 2$ and an opacity at 1\,mm taken from
\citet{Ossenkopf94}.  On the other hand, the mass derived from the
opacity at 1\,mm given in \citet{Ricci10} extrapolated to 3\,mm using
$\beta = 0.9$ would be no greater than 0.3\,\MSUN.

\subsection{Free-free emission} \label{FREE}

\begin{figure}
\begin{tabular}{c}
\includegraphics[width=180mm]{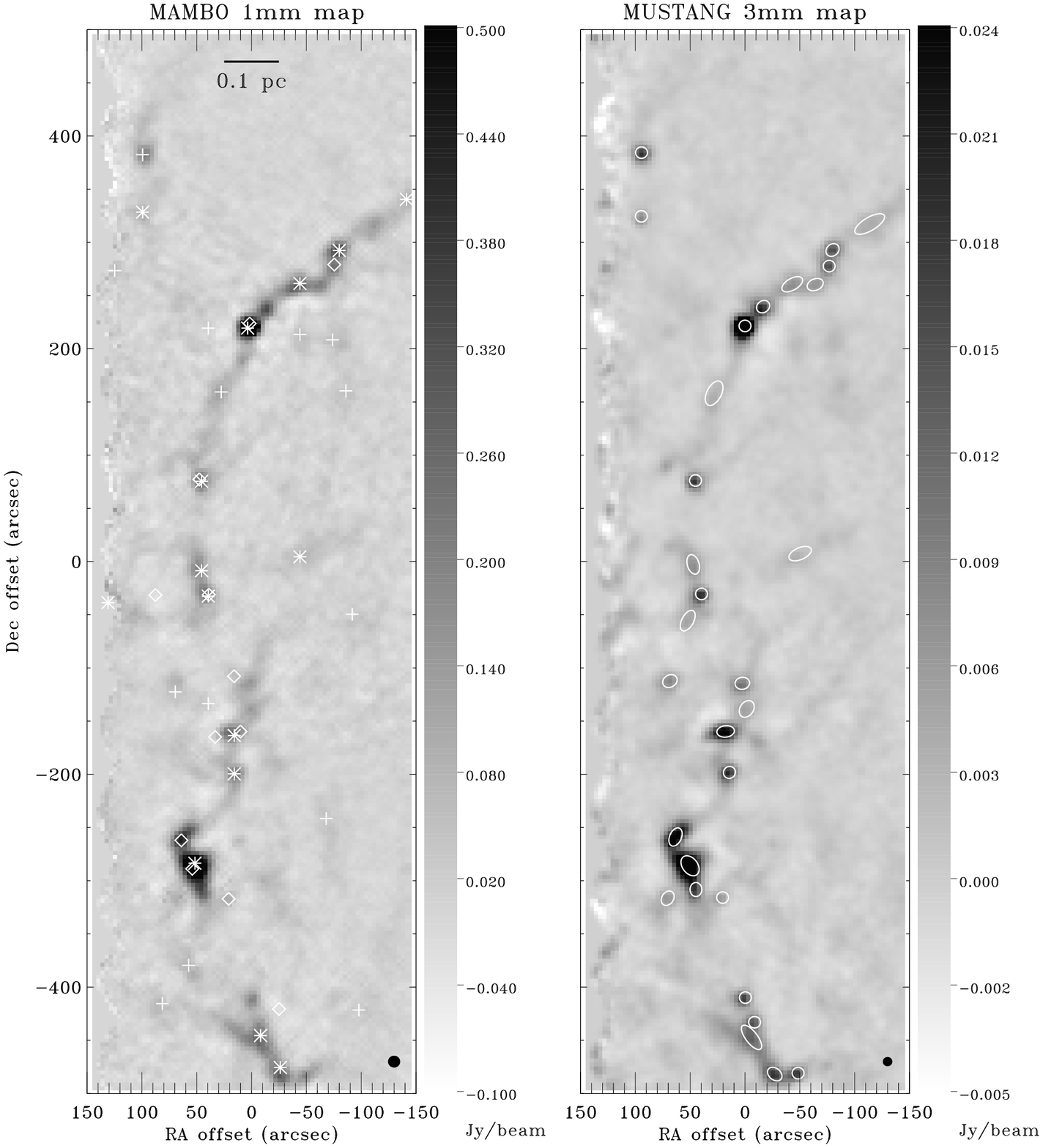}
\end{tabular}
\caption{MAMBO ({\it left}) and MUSTANG ({\it right}) maps of OMC-2/3.
  The MAMBO map has been spatially filtered to recover only those
  scales to which the MUSTANG map is sensitive.  Both maps have units
  of Jy/pixel with 11\arcsec $\times$ 11\arcsec\ pixels. In the 1\,mm
  map, plus signs show the locations of starless cores
  \citep{Sadavoy10}, asterisks show the positions of protostellar
  cores \citep{Sadavoy10}, and diamonds show the locations of 3.6\,cm
  continuum emission detected by \citet{Reipurth99}.  In the map on
  the right, ellipses show the positions of structures identified
  using the {\it getsources} algorithm \citep{Menschikov12}.  The
  (0,0) position is (J2000) 05:35:23.35,
  -05:05:12.5.  \label{FLUXFIG}}
\end{figure}

\begin{figure}
\begin{tabular}{c}
\includegraphics[width=180mm]{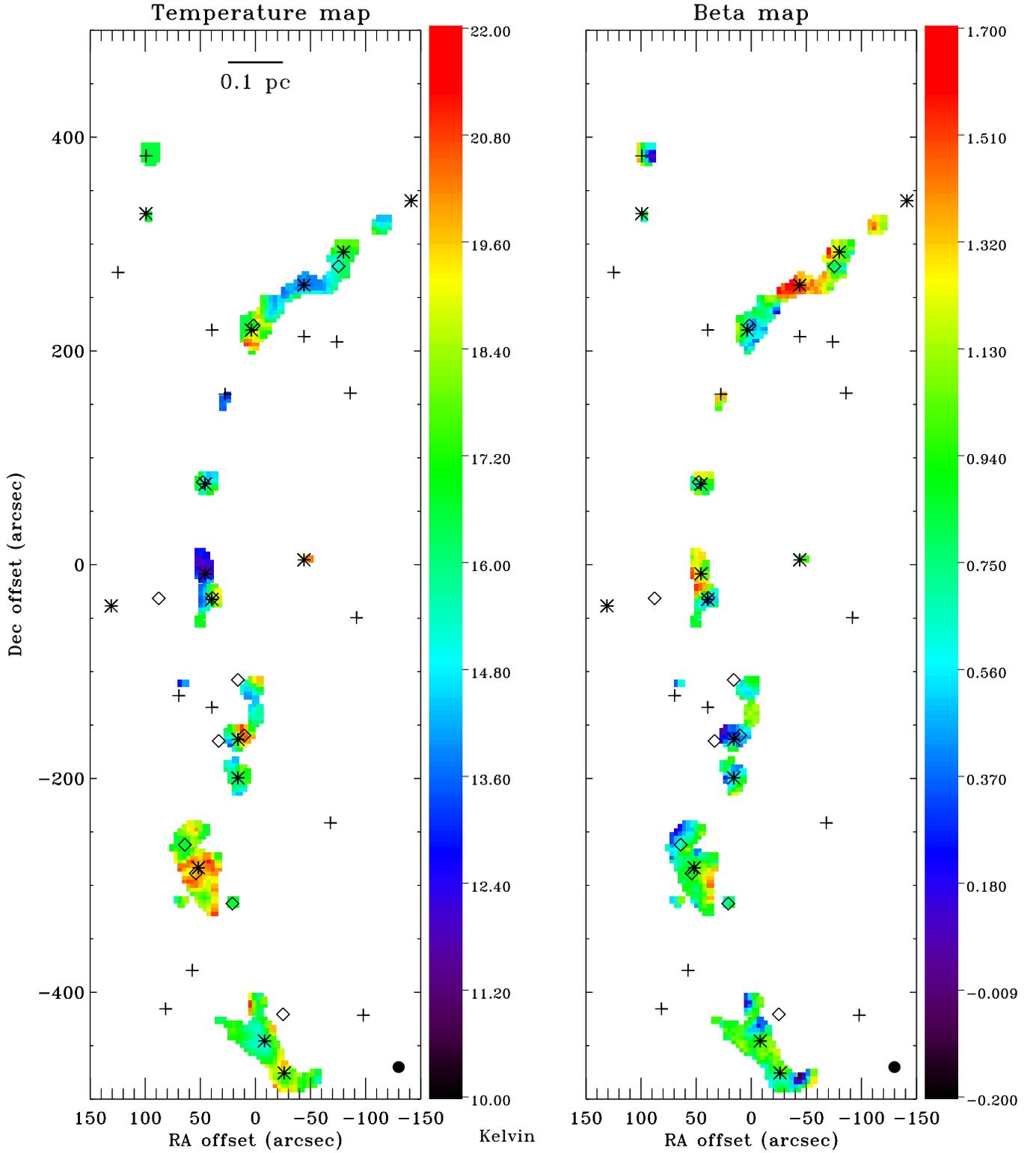}
\end{tabular}
\caption{Temperature ({\it left}) and emissivity spectral index ({\it
    right}) maps of OMC-2/3.  The temperature map comes from
  \citet{Li13} and is smoothed and regridded to the
  11\arcsec\ resolution of the 1\,mm and 3\,mm continuum maps.
  Temperature is only plotted where the emissivity spectral index is
  calculated.  The emissivity spectral index in each pixel is derived
  from the ratio of the 1\,mm and 3\,mm fluxes, using the respective
  temperature map values as the isothermal temperature along the
  line-of-sight.  See Sections \ref{BETAMAP} and \ref{RJ} for details.
  The symbols (plus signs, asterisks, and diamonds) are as in Figure
  \ref{FLUXFIG}.  \label{TEMPBETAFIG}}
\end{figure}

In our determination of $\beta$ (see Section \ref{BETAMAP}), we
assumed that there is no free-free emission contributing significantly
to the observed 1\,mm or 3\,mm maps.  In regions of high-mass star
formation, however, the 3\,mm (and even 1\,mm) continuum can have a
significant contribution from free-free emission.  For instance,
towards Orion BN/KL in OMC-1 (to the immediate south of OMC-2/3), the
3\,mm continuum is evenly split between free-free and thermal dust
emission and in some places the continuum is actually dominated by
free-free emission \citep{Dicker09}.  The OMC-2/3 region lacks the
cluster of OB stars that dominate OMC-1, so we expect the 3\,mm
emission from OMC-2/3 to originate predominantly from dust rather than
ionized gas.  The expectation that the 3\,mm emission is
overwhelmingly from dust is borne out by the similar morphology of the
MUSTANG and spatially filtered MAMBO maps (see Figure \ref{FLUXFIG}).
The assumption that the 3\,mm continuum emission from OMC-2/3 is
purely from dust emission has been made in previous studies
\citep[e.g.,][]{Shimajiri09} and this assumption was also made in the
recent MUSTANG survey of six Class 0 protostars \citep{Shirley11}.

\citet{Reipurth99} used the VLA to map the small-scale 3.6\,cm
emission toward OMC-2/3, finding 14 sources with flux densities
ranging from 0.15\,mJy to 2.84\,mJy at an angular resolution of
8\arcsec.  For optically thin free-free emission ($S_\nu \propto
\nu^{-0.1}$), the contribution of any such emission to the MUSTANG map
would be insignificant, given the $\sim$0.4\,mJy\,beam$^{-1}$ noise in
the map.  For optically thick emission, however, $S_\nu \propto \nu^2$
and free-free emission becomes significant.  We expect some localized
emission at the positions of the OMC-2/3 protostars, which would bias
our derived values of the emissivity spectral index to low values at
these locations.  The free-free emission measured by
\citet{Reipurth99} was unresolved in 11 of the 14 detected sources and
less than 17\arcsec\ in all cases, so any free-free emission from
these sources would be confined to regions one or two resolution
elements wide in our maps.  We suggest that the emissivity spectral
index in regions with $\beta < 0$ is best explained by contamination
from free-free emission toward these objects.  For example, the few
pixels in Figure \ref{TEMPBETAFIG} with $\beta < 0$ generally
correspond to known protostellar cores or the 3.6\,cm continuum
sources (as marked in Figure \ref{FLUXFIG}).  Nevertheless, we do not
expect free-free emission on the spatial scales of filaments to be
significant, and we see low values of $\beta$ associated with such
structures as well.  The starless cores, protostellar cores, and the
filament all have the same average value of $\beta$.  Future
observations of wavelengths between 3.3\,mm and 3.6\,cm would allow us
to determine better the free-free contribution to the 3\,mm map.

\subsection{Anomalous Microwave Emission} \label{AME}

In addition to thermal dust emission and free-free emission, anomalous
microwave emission (AME) can be another important component of the
spectral energy distribution.  A strong candidate for the source of
AME is small, spinning dust grains \citep{Draine12, Draine98}. The AME
contribution to the total SED of molecular clouds has been estimated
in a series of recent papers by the Planck Collaboration.  In the
Gould Belt molecular clouds, the AME typically peaks at
25.5$\pm$0.6\,GHz, falls below thermal dust emission at frequencies
higher than $\sim$30-40\,GHz, and is orders of magnitude smaller than
thermal dust emission at frequencies greater than 80\,GHz, where
MUSTANG is sensitive \citep{PlanckGould}.  In the Perseus molecular
cloud, thermal dust emission dominates AME at frequencies greater than
60\,GHz and the AME contribution at 80\,GHz and above is statistically
insignificant \citep{PlanckAME}.  Similar results are found in the
Milky Way Galactic Plane \citep{PlanckGalactic}, and a wide range of
Galactic clouds \citep{PlanckClouds}.

\begin{figure}
\begin{tabular}{c}
\includegraphics{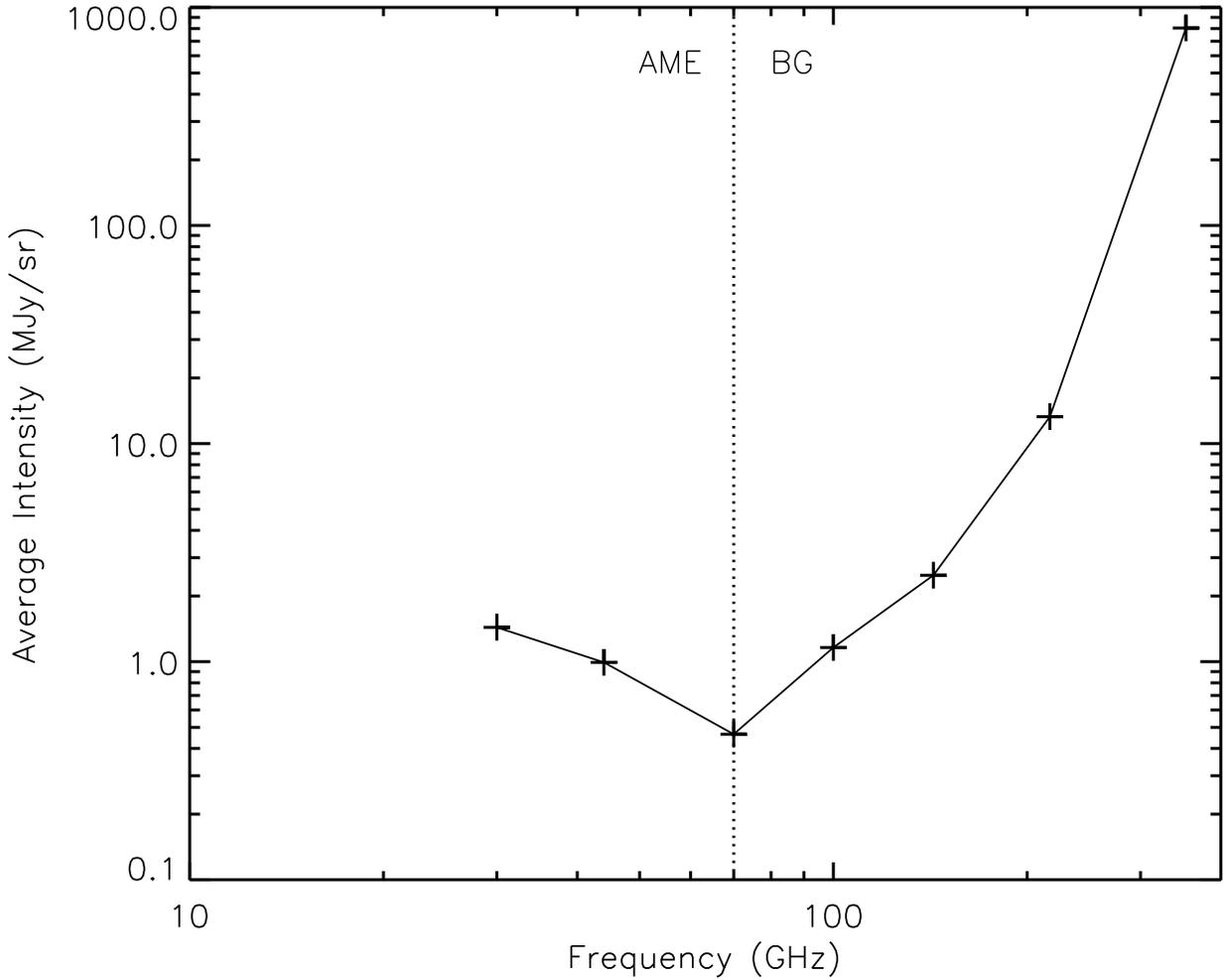}
\end{tabular}
\caption{Planck spectrum of the OMC-3 region, from 30\,GHz to 353\,GHz
  (see Section \ref{AME}). At frequencies less than 70\,GHz, AME
  dominates the SED.  At frequencies greater than 70\,GHz, thermal
  emission from big dust grains (labeled BG here) dominates the
  SED.  \label{PLANCKFIG}}
\end{figure}

To estimate the contribution of AME to the observed 90\,GHz emission,
we use Planck data for OMC-2/3 acquired from the NASA IPAC Infrared
Science Archive.  The Planck resolution at 30\,GHz ($\sim$33\arcmin)
is large enough to blend emission from OMC-1, which at long
wavelengths is dominated by free-free emission, with the emission from
OMC-2.  We therefore restrict our analysis to Planck data between
30\,GHz and 353\,GHz in the OMC-3 region.  We find (see Figure
\ref{PLANCKFIG}) that the emission falls by three orders of magnitude
from 30\,GHz (1\,cm) to 70\,GHz (4\,mm). Therefore, emission at
100\,GHz (3\,mm) and higher frequencies should be dominated by thermal
emission from large dust grains.  We caution that the angular scales
of the Planck maps ($\ge$ 10\arcmin\ at frequencies $\le$ 100\,GHz)
are much larger than the $\sim$10\arcsec\ angular scales probed by our
MUSTANG and MAMBO observations.  Nevertheless, we believe that there
is not much contamination from AME in the MUSTANG 3\,mm map.  Higher
angular resolution maps of 10-70\,GHz continuum emission are needed,
however, to confirm this expectation.

\subsection{CO contamination} \label{CO}

One final possible source of contamination in the 1\,mm and 3\,mm
bands is molecular line emission.  The main source of molecular line
contamination of dust emission is from CO, the brightest molecular
line in regions like OMC-2/3.  For instance, \citet{Drabek12} found
that $^{12}$CO (3-2) emission contributes $\le$20\% of the 850\,$\mu$m
flux observed with SCUBA-2 towards molecular clouds in most regions,
but can be the dominant source of emission towards regions with
molecular outflows.  $^{12}$CO (1-0), at 115\,GHz, falls outside the
range of frequency range (81\,GHz - 99\,GHz) to which MUSTANG is
sensitive \citep{Dicker08}.  $^{12}$CO (2-1), at 230\,GHz, falls
within the MAMBO bandpass of 210\,GHz - 290\,GHz \citep{Bertoldi07}.
Given the temperature and density distributions in OMC-2/3, we would
expect an integrated intensity of CO (2-1) of $\le$
30\,K\,km\,s$^{-1}$ \citep{vanderTak07}.  This would contribute less
than 2\,mJy\,beam$^{-1}$ to our 1\,mm MAMBO map of OMC-2/3, which is
much less than the noise in the map.  An especially bright and broad
(in velocity space) outflow could possibly make a significant
contribution to the 1\,mm map, artificially steepening the derived
value of $\beta$ in its vicinity.  Over the majority of the OMC-2/3
filament, we do not expect molecular line contamination to
significantly affect either the 3\,mm or 1\,mm continuum measurements
of OMC-2/3.

\subsection{Line-of-sight temperature variations} \label{RJ}

In the Rayleigh-Jeans approximation ($h \nu \ll kT$), the observed
flux at a given frequency is linearly dependent on temperature, so the
ratio of fluxes at different frequencies is independent of
temperature.  The Rayleigh-Jeans approximation is only loosely valid
for our observations of the OMC-2/3 region, with $h \nu / k$ values of
12\,K and 4\,K at 250\,GHz and 90\,GHz, respectively.  \citet{Lis98}
found that the temperatures of cores within OMC-2/3 are roughly 17\,K,
with evidence for warmer ($\sim$25\,K) dust in the filament.  The gas
kinetic temperature of the OMC-2/3 filament was measured using GBT and
VLA observations of the NH$_3$ (1,1) and (2,2) inversion transitions
\citep{Li13}.  The gas temperatures are found to be mostly in the
range of 10\,K $\le$ $T_{gas}$ $\le$ 20\,K.  We therefore consider the
possibility that line-of-sight temperature variations influence the
values of $\beta$ determined from the cold dust in OMC-2/3.

\begin{figure}
\begin{tabular}{cc}
\includegraphics[width=80mm]{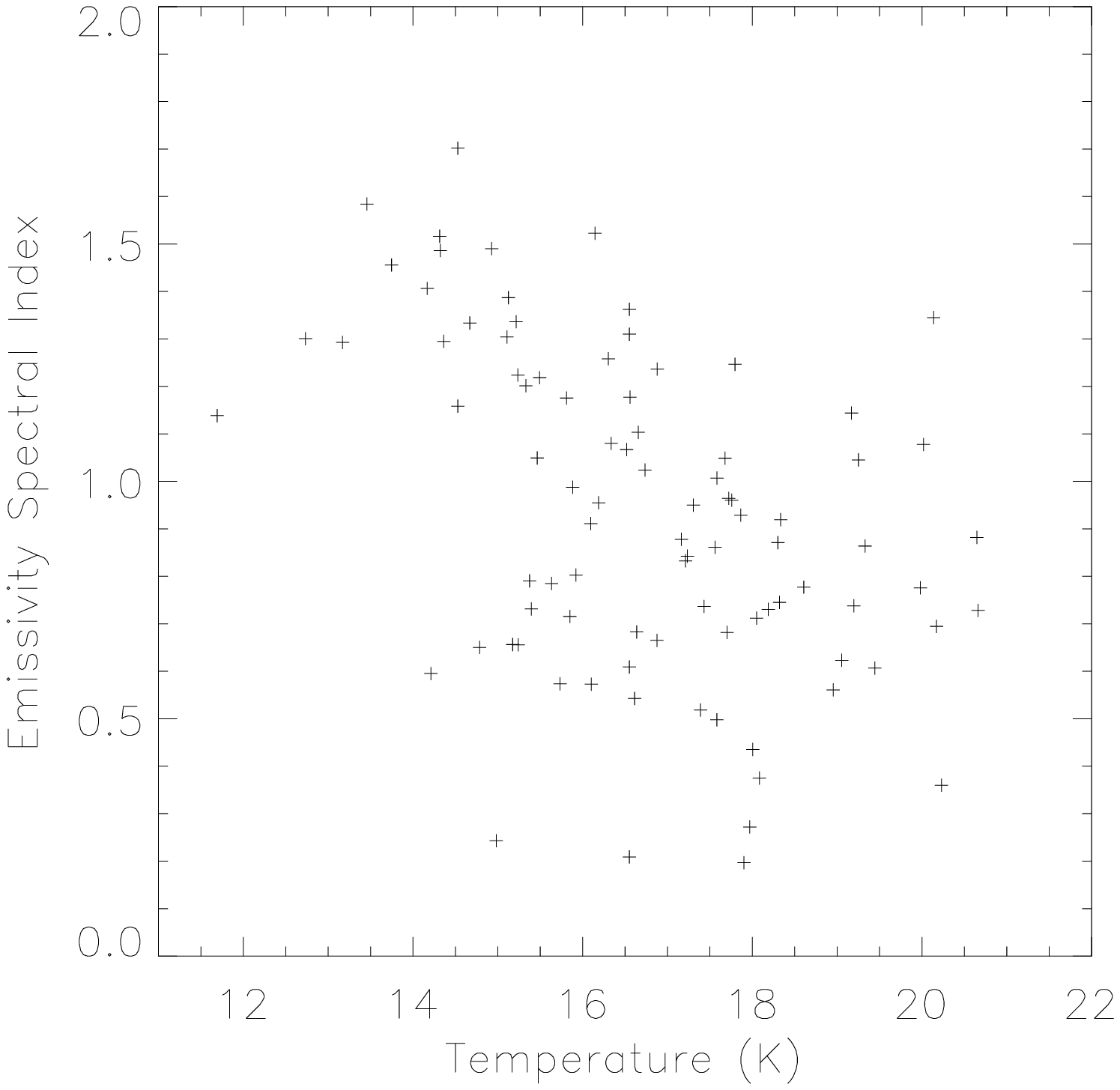} & 
\includegraphics[width=80mm]{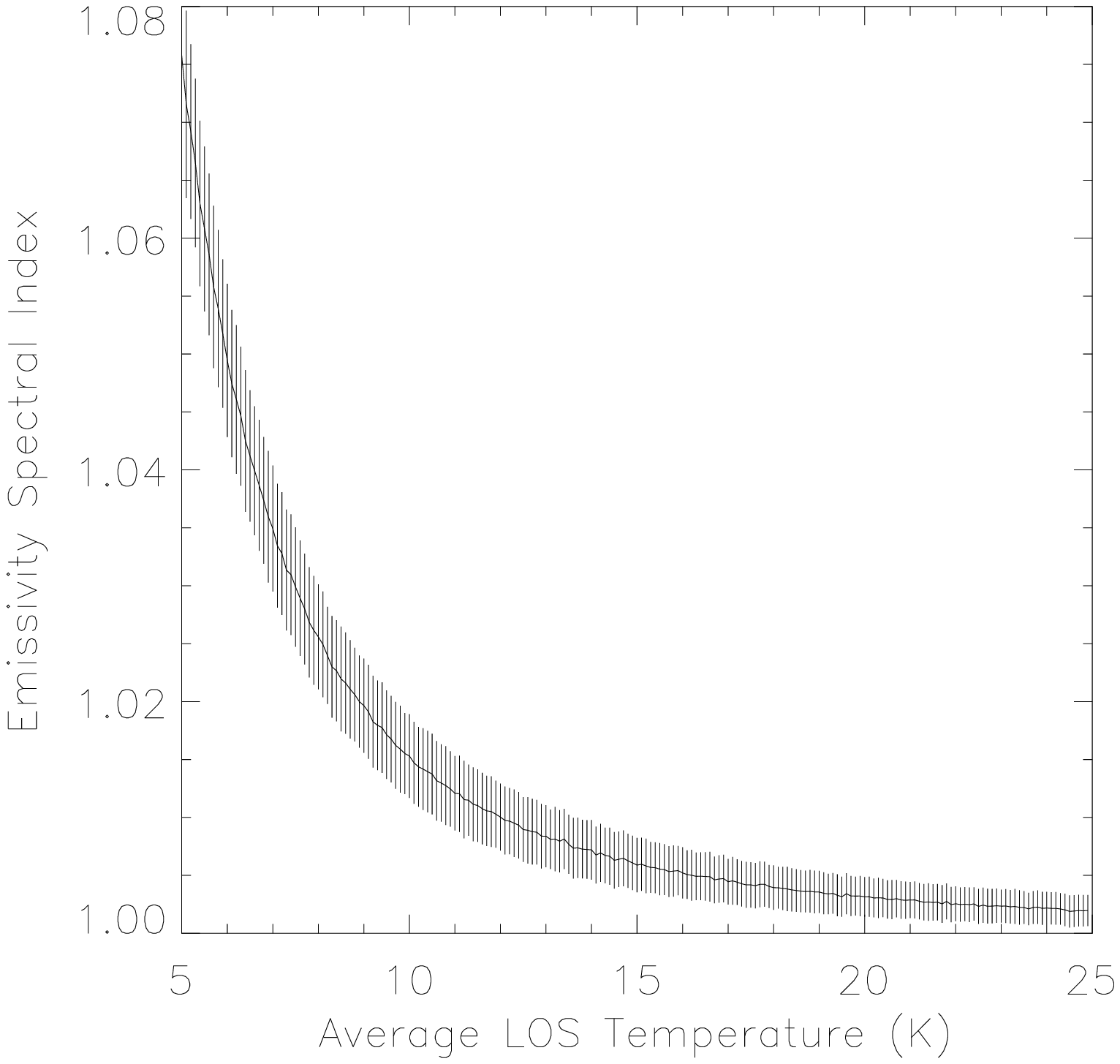} \\
\end{tabular}
\caption{({\it left}) Plot of the emissivity spectral index derived
  from 1\,mm and 3\,mm continuum observations plotted against the
  temperature derived from NH$_3$ (1,1) and (2,2) observations.  Each
  data point is an independent pixel in the maps in Figure
  \ref{TEMPBETAFIG}.  ({\it right}) Results of Monte Carlo simulations
  of the emissivity spectral index that would be derived from 1\,mm
  and 3\,mm flux measurements of dust with a range of temperatures
  along the line of sight, plotted against the average line-of-sight
  temperature.  The dust has an intrinsic emissivity spectral index of
  $\beta = 1$ and the line-of-sight temperature follows a Gaussian
  distribution with a width of 25\% of the mean temperature.  The
  curving black line shows the mean $\beta$ that would be determined
  and the vertical lines show the 1\,$\sigma$ variation around the
  mean.  \label{CORRELATION}}
\end{figure}

When determining the emissivity spectral index map shown in Figure
\ref{TEMPBETAFIG}, we assumed that the temperature along each line of
sight is constant.  The true temperature distribution along each line
of sight is very likely to be variable, so here we estimate the
importance of line-of-sight temperature variations on our $\beta$ map.
We construct a toy model where the dust temperature along a line of
sight is given by a Gaussian distribution with a 1\,$\sigma$ width of
25\% of the mean temperature. We let the mean temperature vary from
5\,K to 25\,K in steps of 0.1\,K.  For each distribution, we calculate
the ratio of the continuum emission at 1.1\,mm and 3\,mm, assuming an
emissivity spectral index of 1.  The 1\,mm/3\,mm flux ratio depends on
the temperature distribution of the dust, especially for low
temperatures where the Rayleigh-Jeans approximation is least valid.
We then use Equation \ref{BETAEQ} to determine the emissivity spectral
index that would be derived from the 1\,mm/3\,mm flux ratio.  We find
that temperature variations lead to a small overestimate of the
emissivity spectral index, i.e., temperature variations along the
line-of-sight push $\beta$ to larger absolute values.  The error is
2\% (or less) for temperature distributions centered on 10\,K or
warmer, and increases to 8\% at a mean temperature of 5\,K (see Figure
\ref{CORRELATION}).  We therefore find that line-of-sight temperature
variations could systematically bias our derived values of $\beta$,
but at a level much lower than the uncertainties introduced by noise
and absolute flux calibration errors.  The systematic bias due to
line-of-sight temperature variations moves $\beta$ in the opposite
direction of our surprising result - i.e., we find that the SED is
shallower than predicted for dust in a filament while line-of-sight
temperature variations work to steepen the SED.

Figure \ref{CORRELATION} compares $\beta$ and temperature for the
entire OMC-2/3 region.  We find that $\beta$ and $T_d$ are
anti-correlated at the 4\,$\sigma$ level.  An anti-correlation between
temperature and the emissivity spectral index has been seen previously
in some molecular clouds \citep[e.g., ][]{Dupac03, Juvela13}, but
other studies have not found conclusive evidence for a
temperature-dependent emissivity spectral index \citep[e.g.,
][]{Veneziani13}.  When using dust emission maps to determine both the
temperature and emissivity spectral index, line-of-sight temperature
variations can create spurious $T_d$-$\beta$ correlations \citep[e.g.,
][]{Shetty09a, Shetty09b, Schnee10, Malinen11, Ysard12, Juvela12a,
  Juvela12b}, though these can be at least partially overcome by
careful analysis and statistical modeling \citep[e.g., ][]{Kelly12,
  Juvela13}.  The analysis here is different from most previous
studies of correlations between temperature and $\beta$ in that these
two quantities are determined from different data sets (NH$_3$
observations and millimeter-wavelength dust emission).  As shown in
Figure \ref{CORRELATION}, this anti-correlation, if real, is not
caused by line-of-sight temperature variations.

\subsection{SED shape} \label{SHAPE}

It is possible that our assumption of a power-law dust emissivity
(i.e., $\kappa_\nu \propto \nu^\beta$) between 90\,GHz and 250\,GHz is
not justified.  For certain size distributions and compositions,
detailed models of the dust opacity between 90\,GHz and 250\,GHz show
that it has a more complicated form \citep[e.g.,][]{Draine06}.
Laboratory measurements and theoretical models of dust grain analogues
have found that the emissivity can be temperature-dependent
\citep[e.g.,][]{Agladze96, Boudet05, Meny07} and not well described by
a simple power-law \citep[e.g.,][]{Coupeaud11, Paradis11}.  The
relatively low values of $\beta$ we find might not be a result of
large dust grains, but rather may be determined by the grain
composition, temperature, and the wavelengths of our observations.
Observations of circumstellar disks, however, often sample the SED
from sub-millimeter to centimeter wavelengths and these are
successfully modeled with a simple power-law opacity
\citep[e.g.,][]{Lommen07, Perez12}.

\subsection{Calibration errors} \label{CALIBRATION}

A trivial explanation for the low values of $\beta$ found in OMC-2/3
would calibration errors.  Given the relatively large frequency range
between the MAMBO and MUSTANG observations, it would require a large
error in the flux calibration to bring the emissivity spectral index
into agreement with values found in Perseus B1 or the Orion Nebula
Cluster.  For instance, if we decrease the fluxes in the 3\,mm map by
10\% and increase the fluxes in the 1\,mm map by 10\%, then the median
value of $\beta$ increases from 0.9 to 1.0 (the standard deviation of
derived values of $\beta$ remains 0.3).  To bring the median value of
$\beta$ up to 2, we would need to increase the ratio of $S_{\rm
  250\,GHz}/S_{\rm 90\,GHz}$ by a factor of 4.  We do not believe that
the flux uncertainties in the MAMBO or MUSTANG maps are anywhere near
that high, so the absolute calibration of the bolometer maps is not
likely to be the main driver of the low values of $\beta$ found here.

\subsection{Interpretation} \label{Calibration}

Of the possible explanations for the shallow emissivity spectral index
in OMC-2/3, we find that the two most likely are the presence of large
grains or a dust emissivity not well described by a single power-law.
If dust grains in OMC-2/3 are characterized by millimeter size scales,
this would be the first report of such large grains in structures with
scales on the order of 1\,pc.  The protostellar cores and
circumstellar disks with millimeter-sized grains have sizes of $\loa$
0.1\,pc.  It could be important to models of grain growth in disks and
cores if the dust in filaments is already quite large.  In this case,
it will be important to determine if OMC-2/3 is unique in exhibiting
large grains or if this is a common feature of star-forming
filaments.  Although OMC-2/3 is unique in that it has a higher density
of starless and protostellar cores than other regions within
$\sim$500\,pc of the Sun, there are no other properties (mass,
density, temperature, etc.) that would lead one to suspect that the
dust grains in OMC-2/3 ought to have properties significantly
different than those found in other nearby molecular clouds.

Alternatively, the shallow spectral index between 1\,mm and 3\,mm
found in OMC-2/3 is due to an SED that is not well-characterized by a
power law function at millimeter wavelengths.  In this case, the
grains are not necessarily larger than is commonly found in nearby
molecular clouds.  It would be important to take the true shape of the
dust emissivity into account when conducting future studies of $\beta$
in filaments.  In either case, more detailed observations are needed
to study the full dust SED in filaments to make accurate measurements
of mass, temperature, and possible large grains.

\section{Future Work}

Some of the assumptions made in Section \ref{ANALYSIS} and some of
the potential sources of error given in Section \ref{DISCUSSION} can
be tested with additional observations.  

We have made the assumption that all of the 3\,mm flux comes from dust
emission, but it seems clear that free-free emission may contribute
somewhat to the observed 3\,mm map, especially near protostars.  The
contribution of free-free emission to the 3\,mm map can be tested by
making observations at wavelengths between the 3.6\,cm map produced by
\citet{Reipurth99} and the 3.3\,mm map shown here.  These observations
could be used to determine at what wavelength the free-free emission
observed by \citet{Reipurth99} switches from being optically thick to
thin, allowing us to subtract accurately the free-free component from
the 3\,mm map and make a more accurate map of the emissivity spectral
index.

We have made the assumption that the dust opacity is given by a power
law, but this assumption can be tested by additional observations.  To
avoid uncertainties introduced by line-of-sight temperature
variations, observations at 850\,$\mu$m (with SCUBA-2, for example) or
2\,mm (with the IRAM 30\,m, for example) could be used to constrain
better the shape of the (sub)millimeter SED and minimize the effect of
uncertainties in absolute flux calibration on the derived value of the
emissivity spectral index.  Herschel observations between
70\,\micron\ and 500\,\micron\ could also be folded into this
analysis, with similar spatial filtering and careful corrections for
temperature variations and optical depth.  High-resolution extinction
maps could also be used to constrain the dust opacity independently
from the dust temperature distribution.

\section{Summary} \label{SUMMARY}

We have mapped the emissivity spectral index of the OMC-2/3 using new
3\,mm observations paired with previously published observations of
the 1\,mm continuum and NH$_3$-derived gas temperature.  Focusing more
on spatial scales of filaments and cores, we find that the emissivity
spectral index is much shallower than is often assumed for dust in
molecular clouds, with $\beta = 0.9 \pm 0.3$.  We find a weak
correlation between $\beta$ and $T_d$, though there are no significant
differences between the average values of $\beta$ found in the OMC-2/3
filament, in the starless cores, and in the protostellar cores.  Such
a low average value of $\beta$ in OMC-2/3 could be explained by
millimeter-sized dust grains, as inferred in many circumstellar disks.
Dust emissivities that vary with temperature or opacities that do not
vary with wavelength as a simple power law could also explain our
observations.  We do not expect either free-free emission or AME to be
significant on the scale of the 2 parsec-long OMC-2/3 filament,
though, if present and sufficiently bright, either emission mechanism
would result in lower absolute values of $\beta$.  We suggest future
observations that can be used to determine the strength of free-free
emission and AME.

\section*{Acknowledgments}

We thank our anonymous referee for comments that improved the content
and clarity of this paper.  The National Radio Astronomy Observatory
is a facility of the National Science Foundation operated under
cooperative agreement by Associated Universities, Inc.  JDF
acknowledges support by the National Research Council of Canada and
the Natural Sciences and Engineering Council of Canada(NSERC) via a
Discovery Grant.  RF is a Dunlap Fellow at the Dunlap Institute for
Astronomy and Astrophysics, University of Toronto. The Dunlap
Institute is funded through an endowment established by the David
Dunlap family and the University of Toronto.  DL is supported by China
Ministry of Science and Technology under State Key Development Program
for Basic Research (2012CB821800).  The authors would like to thank
the MUSTANG instrument team from the University of Pennsylvania, NRAO,
Cardiff University, NASA-GSFC, and NIST for their efforts on the
instrument and software that have made this work possible.  D. Li
acknowledges the support from National Basic Research Program of China
(973 program) No. 2012CB821800, NSFC No. 11373038, and Chinese Academy
of Sciences Grant No. XDB09000000.

{\it Facilities}: GBT, VLA, IRAM 30\,m

{}

\clearpage

\clearpage

\clearpage

\clearpage
\begin{figure}
\begin{tabular}{c}
\includegraphics{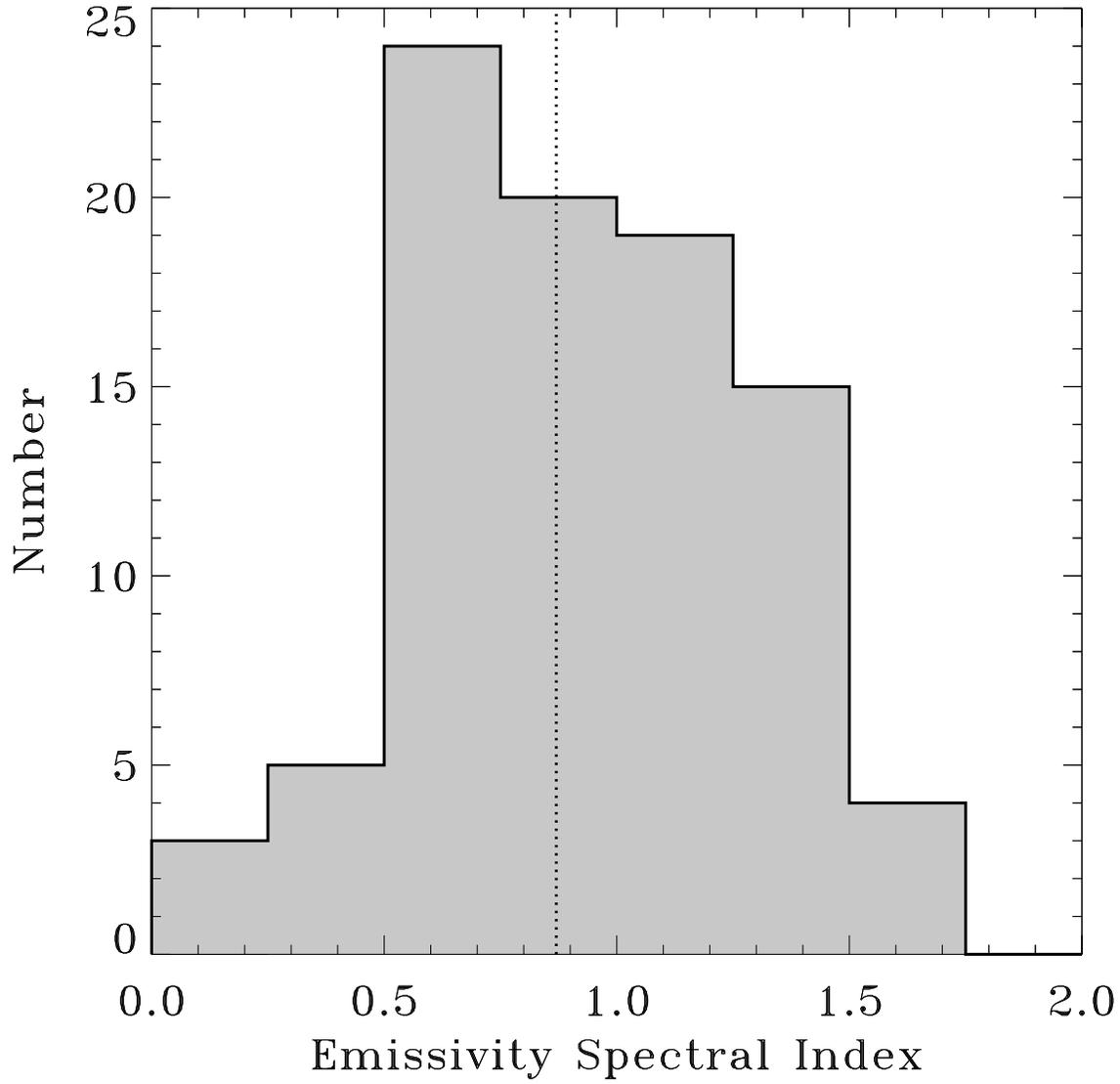}
\end{tabular}
\caption{Histogram of the emissivity spectral indices measured in
  OMC-2/3.  The vertical dashed line shows the median value of $\beta$.
  See Sections \ref{BETAMAP} for details.  \label{BETAHIST}}
\end{figure}

\clearpage

\clearpage
\begin{deluxetable}{ccccccc} 
\tablewidth{0pt}
\tabletypesize{\scriptsize}
\tablecaption{High significance {\it getsource} objects \label{SOURCETABLE}}
\tablehead{
 \colhead{RA}     &
 \colhead{Dec}     &
 \colhead{Major Axis\tablenotemark{1}}      &
 \colhead{Minor Axis\tablenotemark{1}}      &
 \colhead{Position Angle\tablenotemark{1}}  &
 \colhead{Peak Flux\tablenotemark{1}}       &
 \colhead{Integrated Flux\tablenotemark{1}} \\
 \colhead{J2000}                            &
 \colhead{J2000}                            &
 \colhead{FWHM}                             &
 \colhead{FWHM}                             &
 \colhead{}                                 &
 \colhead{}                                 & 
 \colhead{}                                 \\
 \colhead{decimal $^\circ$}                   &
 \colhead{decimal $^\circ$}                   &
 \colhead{\arcsec}                          &
 \colhead{\arcsec}                          &
 \colhead{$^\circ$ E of N}                    &
 \colhead{Jy/beam}                          & 
 \colhead{Jy}}
\startdata
83.8473 & -5.0253 & 10.8 & 10.8 & 161.9 & 0.112 & 0.146 \\
83.8613 & -5.1662 & 20.8 & 14.0 &  35.2 & 0.036 & 0.109 \\
83.8650 & -5.1587 & 18.0 & 10.8 & 157.6 & 0.028 & 0.053 \\
83.8522 & -5.1311 & 16.0 & 10.8 &  96.9 & 0.026 & 0.050 \\
83.8584 & -5.0952 & 10.8 & 10.8 & 140.3 & 0.020 & 0.027 \\
83.8599 & -5.0656 & 10.8 & 10.8 &  28.4 & 0.018 & 0.024 \\
83.8427 & -5.0202 & 12.8 & 10.8 & 129.8 & 0.020 & 0.033 \\
83.8513 & -5.1419 & 11.4 & 10.8 & 149.2 & 0.018 & 0.026 \\
83.8250 & -5.0054 & 13.1 & 10.8 & 134.5 & 0.014 & 0.025 \\
83.8472 & -5.2006 & 10.8 & 10.8 & 103.5 & 0.017 & 0.020 \\
83.8338 & -5.2204 & 10.8 & 10.8 & 110.5 & 0.018 & 0.020 \\
83.8736 & -4.9800 & 10.8 & 10.8 & 100.9 & 0.018 & 0.029 \\
83.8397 & -5.2206 & 14.9 & 10.8 &  48.4 & 0.016 & 0.030 \\
83.8259 & -5.0096 & 10.8 & 10.8 & 178.1 & 0.015 & 0.018 \\
83.8598 & -5.1724 & 13.2 & 10.8 & 179.5 & 0.015 & 0.022 \\
83.8353 & -5.0144 & 21.1 & 10.9 & 120.9 & 0.007 & 0.017 \\
83.8449 & -5.2071 & 10.8 & 10.8 & 160.0 & 0.012 & 0.015 \\
83.8480 & -5.1186 & 13.3 & 11.3 & 100.1 & 0.010 & 0.019 \\
83.8456 & -5.2110 & 27.4 & 10.8 &  36.5 & 0.009 & 0.022 \\
83.8605 & -5.0875 & 18.8 & 10.8 &  15.8 & 0.007 & 0.015 \\
83.8530 & -5.1745 & 10.8 & 10.8 & 105.4 & 0.006 & 0.006 \\
83.8736 & -4.9966 & 10.8 & 10.8 & 173.4 & 0.008 & 0.010 \\
83.8294 & -5.0145 & 15.3 & 10.8 & 114.2 & 0.006 & 0.012 \\
83.8664 & -5.1180 & 13.9 & 11.3 & 122.7 & 0.008 & 0.018 \\
83.8670 & -5.1747 & 14.4 & 10.8 & 149.4 & 0.006 & 0.009 \\
83.8469 & -5.1253 & 16.8 & 12.3 & 148.5 & 0.004 & 0.009 \\
83.8619 & -5.1022 & 21.1 & 10.8 & 152.8 & 0.005 & 0.008 \\
83.8156 & -4.9987 & 31.1 & 12.3 & 121.8 & 0.003 & 0.010 \\
83.8552 & -5.0428 & 24.7 & 13.3 & 154.0 & 0.004 & 0.012 \\
83.8333 & -5.0847 & 21.9 & 11.1 & 114.4 & 0.003 & 0.007 \\
\enddata
\tablenotetext{1}{From 3\,mm flux map}
\end{deluxetable}

\end{document}